\documentclass[twocolumn,showkeywords,preprintnumbers,amsmath,amssymb]{revtex4}

\usepackage{graphicx}
\usepackage{dcolumn}
\usepackage{bm}

\begin{document}

\title{Universal Grammar and Consciousness}

\author{Daegene Song}

\affiliation{%
Department of Management Information Systems, Chungbuk National University, Cheongju, Chungbuk 28644, Korea
}%

\date{\today}

\begin{abstract}

The orthodox interpretation of quantum theory treats the subject and the object on an equal footing. It has been suggested that the cyclical-time process, which resolves self-reference in consciousness, interconnects the observed universe and the mind of the subject. Based on the analogy between cryptography and language, the concept of the common innate structure of language, also known as universal grammar, may be associated with the continuity in consciousness. Extending this connection, L\'{e}vi-Strauss's proposal on universal culture may be considered as a shared structure of continuity among the consciousness of multiple subjects.

\end{abstract}

\maketitle

\section{Introduction}
 
 In his theory of Forms, Plato argued that the continuously changing physical world may not be reliable. Instead, he maintained that there must be a world with idealistic and unchanging forms, such as a perfect right triangle or a perfect circle. In particular, he viewed this imperfect physical world as a representation of the perfect world. Plato argued that it is important to understand this ideal world in order to obtain useful knowledge. This process often happens in modern science where theoretical modeling is often done based on ideal situations, which may approximate the actual imperfect phenomena.

While there are a variety of ideas involving the interpretation of quantum theory, most physicists agree that the standard quantum theory provides a precise description of what happens. Nevertheless, the quantum world is different from the physical world and resides in imaginary space composed of imaginary numbers (Figure 1). Werner Heisenberg succinctly put this into {\it{I think that modern physics has definitely decided in favor of Plato}} (Heisenberg, 1981).  Moreover, Heisenberg went on to state,
\begin{center} 
{\it{In fact the smallest units of matter are not physical objects in the ordinary sense; they are forms, ideas which can be expressed unambiguously only in mathematical language.}}
\end{center}
For instance, why is it possible for humans to easily imagine perfect or ideal forms when nobody has ever seen any? A similar case can be found with infinite real numbers that exist in the line connecting the natural numbers, such as 0 and 1 - that is, continuous real numbers are not observable in a physical space. The continuity exists only in thought.


\begin{figure}
\begin{center}
\includegraphics[width=0.45\textwidth]{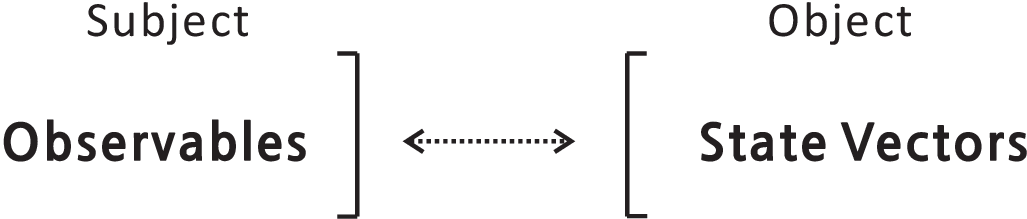}

\end{center}
\caption{ The Copenhagen Interpretation of quantum theory taken to the limit implies the object (i.e., the whole observed universe) to be represented by the state vector and the consciousness of the subject with observables.      }

\end{figure}

On the other hand, Niels Bohr discussed physics to be associated with a posteriori type of knowledge associated with an empirical experience rather than an a priori one. However, he also tried to outline how objectivity may also be related to human language, as in the following (Bohr, 1960),
\begin{center}
{\it{In this respect our task must be to account for such experience in a manner independent of individual subjective judgement and therefore objective in the sense that it can be unambiguously communicated in ordinary human language.}}
\end{center}


\section{Philosophical Thoughts}
Karl Popper, a philosopher of science, also mentioned the limited or subjective access science could provide. Indeed, Popper pointed out that science is developed based on moving from observation to developing theory. Popper described an episode with his students in Vienna in which the students had difficulty to simply write down what they observed. Thus, for Popper, observation is subjective and depends on personal background, tendency, and interests. 

The twentieth-century analytic philosopher Ludwig Wittgenstein presented a theory that compares the picture with reality where both the picture and reality are composed of individual elements such that they share the same logical structure: {\it{The logical structure of the picture, whether in thought or in language, is isomorphic with the logical structure of the state of affairs which it pictures}} (Wittgenstein, 1922).

As Wittgenstein discussed, language and the world may share the same logical structure. Popper also pointed out the role that language plays in this selective observation such that there exist various presuppositions in language and consciousness which may depend on personal history, social interests, or genetic tendency. Describing the fundamental limit of scientific reasoning, Popper described the following: {\it{Science may be described as the art of systematic oversimplification}} (Popper, 1992).

Gottlob Frege, the founder of modern logic, was interested in finding a structure that is independent of psychology or human thought, as the following quote of his indicates, {\it{Being true is different from being taken to be true, …  I understand by \lq laws of logic\rq not psychological laws of takings-to-be-true …}} (Frege, 1964).  Indeed, Frege wanted to develop logical laws for truth as {\it{stones set in an eternal foundation}} that human thought could not replace.   

In his book {\it{The Critique of Pure Reason}}, Immanuel Kant explained the combination of empiricism and rationalism as follows: {\it{All our knowledge begins with the senses, proceeds then to the understanding, and ends with reason}}.  On the other hand, Karl Popper was critical of the inductive reasoning employed in science. Instead, he viewed science as the continuous effort to get to the truth through testing (Popper, 1972): {\it{Induction is logically invalid; but refutation or falsification is a logically valid way of arguing from a single counterinstance to \lq or, rather, against\rq the corresponding law.}}


\begin{figure}
\begin{center}
\includegraphics[width=0.4\textwidth]{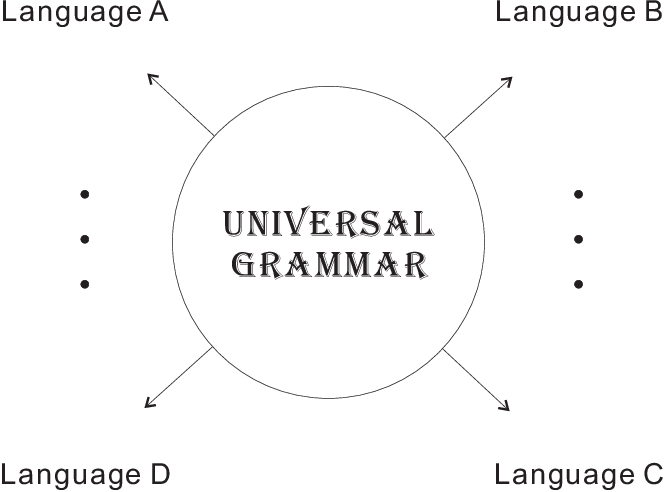}

\end{center}
\caption{ Universal grammar suggests different languages share a common and an innate structure.   }

\end{figure}

\section{Language and Culture}
Anyone who has tried to learn a foreign language understands how difficult it is, and almost always, the second language does not become as comfortable as the first native language. Why is this the case? The difficulty is only magnified when one attempts to learn the second language by understanding its sophisticated grammatical rules. On the other hand, children who learn the language at early age acquire the capacity to speak a native language. The more puzzling part is that this process of native language acquisition is often done without learning any grammar of the language. The linguist Noam Chomsky made the bold proposal that one is born with a capacity to speak a language, which became known as the universal grammar (Chomsky, 1959; 1980). Indeed, in the latter part of the twentieth-century, the method of generative grammar prevailed. In particular, Chomsky expanded and analyzed that all human language may have a common grammatical structure (Figure 2).


\begin{figure}
\begin{center}
\includegraphics[width=0.4\textwidth]{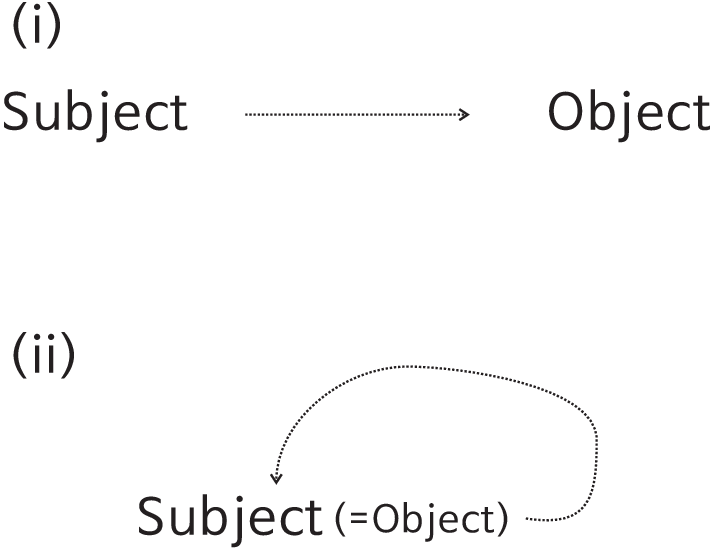}

\end{center}
\caption{ (i) Ordinary case where the observer is observing the object. (ii) In the case of self-referential consciousness, the object being observed is the subject.     }

\end{figure}

In (Song, 2017a), the interconnection between mind and matter were proposed based on the cyclical-time approach. Although the usual phenomenon corresponds to the subject observing the object (Fig 3 (i)), in the case of self-referential consciousness, the object is also the subject (Fig 3 (ii)). In order to resolve the paradox resulting from self-reference, the cyclical-time model was employed to explain the apparent discrepancy between the physical and quantum vacuums associated with the cosmological constant problem (Figure 4). Since one of the motivations for introducing a new model of the universe was the question surrounding the fundamental nature of humans (Song, 2017a; 2017b), it is reasonable to ask if the new model can explain the special phenomena involving language. 

Indeed, the new model suggested the discrete classical phenomena are linked with continuous semantics. This resembles the surprising aspect of language in the sense that finite and discrete symbols can carry infinite, ideal, or continuous meanings. To solve this strange situation, the subjects ought to share continuity or infinity beforehand - similar to shared secret keys in cryptography (Song, 2018). Moreover, this innate, or pre-shared, aspect of language should correspond to the universal grammar. One of the suggestions involving the structural aspect of universal grammar is the recursive aspect, which is similar to the liar's paradox:
\begin{itemize}
\item	This sentence is false.
\item	This sentence (which refers to \lq this sentence is false\rq) is false.
\item	This sentence (which refers to \lq this sentence [which refers to \lq this sentence is false\rq ] is false\rq ) is false.
\end{itemize}
The capacity to generate a sentence with a recursive nature is seen in natural language. This is equivalent to the following self-referential aspect of consciousness:
\begin{itemize}
\item	The observer observes the observer.
\item	The observer observes the observer (who observes the observer).
\item	The observer observes the observer (who observes the observer [who observes the observer]).
\end{itemize}


\begin{figure}
\begin{center}
\includegraphics[width=0.4\textwidth]{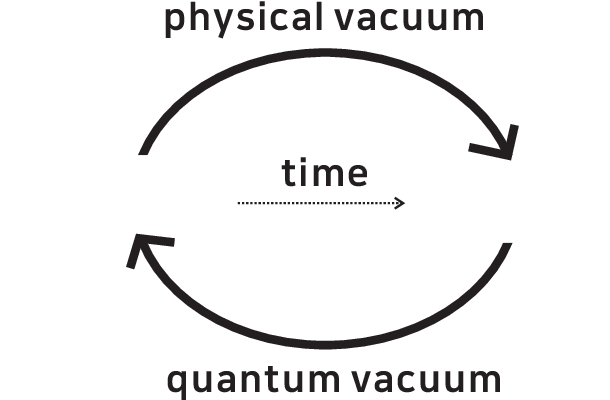}

\end{center}
\caption{ The cyclical-time model of the universe suggests the time forward physical vacuum is filled with the subject's conscious quantum vacuum, which is going backwards in time.  }

\end{figure}

In anthropology, structuralism was developed by the French scholar Claude L\'{e}vi-Strauss to analyze human culture in terms of its structural connections. In particular, L\'{e}vi-Strauss' approach was that there are sophisticated universal structures present in any human culture system (L\'{e}vi-Strauss, 1955). He particularly attempted to associate innate aspects of the human mind with this universal cultural pattern. The argument connecting the subject model of interwoven matter and mind through cyclical time with universal grammar in linguistics may be extended to consider multiple subjects (Figure 5). That is, the continuity of consciousness in the universe model may be associated with the structure of universal culture as discussed by L\'{e}vi-Strauss. The existentialist philosopher Husserl also discussed the interconnectivity between different consciousness, which may enter as empathy.


\begin{figure}
\begin{center}
\includegraphics[width=0.4\textwidth]{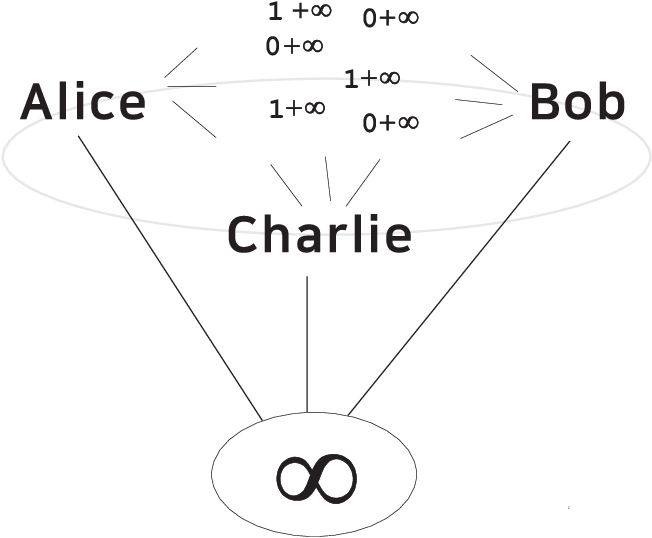}

\end{center}
\caption{ The cryptographic analysis of language suggests that people share a common capacity to understand continuity, or infinity, while discrete and finite languages are exchanged.      }

\end{figure}

\section{Pioneers}
The inseparability between the subject and the object, or matter and mind, has also been considered in science. In particular, two prominent physicists presented similar bold conjectures. The first, David Bohm, was born in 1917 in USA, the son of Jewish immigrants. Bohm obtained his doctorate from the University of California at Berkeley and started to teach at Princeton University. However, due to his involvement in a radical political movement, he was unable to continue to work at the university. He moved to Brazil and became a physics professor at the University of Sao Paulo and eventually settled at Berbeck College in London. 

In his book {\it{Wholeness and Implicate Order}}, Bohm emphasized the inseparable wholeness. Indeed, Bohm discussed a similar idea to the proposal of the inseparability of subject and object in terms of {\it{universal flux}}, which is considered as an essential element that unites mind and matter (Bohm, 1980): 
\begin{center}
{\it{ … a universal flux that cannot be defined explicitly but which can be known only implicitly …  In this flow, mind and matter are not separate substances.}} 
\end{center}

The second scientist and one of the important pioneers in the field of physics was John Archibold Wheeler, who was born in 1911. Not only was he a great researcher with deep intuition, but he also had numerous students who later became well-known physicists, including the Nobel laureate Richard Feynman and Hugh Everett, who proposed many world interpretations of quantum theory. Wheeler obtained his doctorate from John Hopkins University and also participated in the Manhattan Project during World War II. He proposed the concept of a participatory universe and it from bit that the existence of physical reality derives from information: {\it{It from bit symbolizes the idea that every item of the physical world as at bottom … an immaterial source …}} (Wheeler, 1990). 

In particular, three concepts within Wheeler's discussion in developing the new insight into the structure of the universe bear resemblance to the proposal the of cyclical-time subject model. First, Wheeler discussed the importance of considering the existence of the universe and the observer with consciousness—that is, the very existence of physical reality may be related to observership. It has been argued that due to the paradox resulting from the self-referential consciousness, the subject and the object are inseparable (Song, 2007; 2017a).
 
Second, with his well-known phrase {\it{it from bit}}, Wheeler outlined his idea about the connection between the physical world and the metaphysical elements of an immaterial source and explanation, which would be connected through information, and he called it the {\it{participatory universe}} (Wheeler, 1990). Indeed, the proposal in (Song, 2017b) discussed how information, or language, plays a central role in connecting the physical reality and consciousness. 

Thirdly, Wheeler noted the importance of time in explaining existence itself where he said {\it{Of all obstacles to a thoroughly penetrating account of existence, none looms up more dismayingly than \lq time\rq }} (Wheeler, 1986).  As Wheeler pointed out, time, particularly cyclical time, is an essential element in describing a new perspective of the universe, as indicated in (Song, 2017a).   

\begin{figure}
\begin{center}
\includegraphics[width=0.4\textwidth]{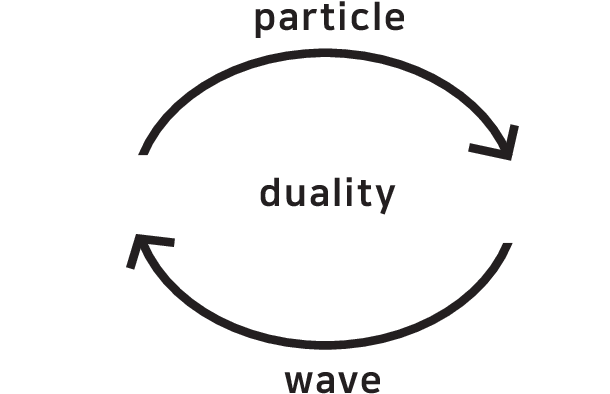}

\end{center}
\caption{The wave-particle duality may also be considered in terms of the cyclical-time process interweaving matter (particle) and mental (wave) parts.       }

\end{figure}

\section{Remarks}
Karl Popper outlined both the strengths and the weaknesses of science. Although science is often based on oversimplifying induction, its tendency to previous errors allows it to continually make progress.  

Notable physicists, such as Wheeler and Bohm, have suggested the radical idea that the subject and the object may not be separable. Indeed, this inseparability may correspond to the physical universe being filled with the observer's consciousness through cyclical time. In fact, one of the puzzling aspects of quantum theory is that the wave-particle duality may be understood better in the new model. While the particle corresponds to the physical matter in a time-forward manner, the wave aspect corresponds to the subject's conscious awareness that is going backwards in time (Figure 6).

In this paper, it was discussed that the continuity in the subject's consciousness may be a shared one, which would correspond to universal grammar in linguistics and to universal culture, as proposed by L\'{e}vi-Strauss. This also explains how people can better communicate concepts involving continuity or infinity with only discrete and finite bits.


\end{document}